\documentclass[11pt,preprintnumbers,titlepage,nofootinbib]{revtex4-2}
\usepackage[utf8]{inputenc}
\usepackage{amsmath}
\usepackage{amssymb}
\usepackage{graphicx}
 \usepackage[bookmarks=false,
 breaklinks=false,pdfborder={0 0 1},colorlinks=false]
 {hyperref}
\hypersetup{
 colorlinks,linkcolor=blue,citecolor=blue,urlcolor=blue}

\makeatletter

\DeclareTextSymbolDefault{\textquotedbl}{T1}

\newenvironment{lyxlist}[1]
	{\begin{list}{}
		{\settowidth{\labelwidth}{#1}
		 \setlength{\leftmargin}{\labelwidth}
		 \addtolength{\leftmargin}{\labelsep}
		 }}
	{\end{list}}

\usepackage{wasysym}
\usepackage{array}
\usepackage{multirow}
\usepackage{wasysym}
\usepackage{wasysym}
\usepackage{amsfonts}
\usepackage{subfigure}
\usepackage{cleveref}
\usepackage{listings}
\usepackage{xcolor}
\usepackage{array}
\usepackage{url}
\usepackage{color}

\@ifundefined{textcolor}{}{
\definecolor{BLACK}{gray}{0}
\definecolor{WHITE}{gray}{1}
\definecolor{RED}{rgb}{1,0,0}
\definecolor{GREEN}{rgb}{0,1,0}
\definecolor{BLUE}{rgb}{0,0,1}
\definecolor{CYAN}{cmyk}{1,0,0,0}
\definecolor{MAGENTA}{cmyk}{0,1,0,0}
\definecolor{YELLOW}{cmyk}{0,0,1,0}
}

\makeatother

\begin{document}
\title{Stable long-term evolution in numerical relativity}

\author{Sebastian Garcia-Saenz$^{a}$}
\email{sgarciasaenz@sustech.edu.cn}

\author{Guangzhou Guo$^{a}$}
\email{guogz@sustech.edu.cn}

\author{Peng Wang$^{b}$}
\email{pengw@scu.edu.cn}

\author{Xinmiao Wang$^{a}$}
\email{12132943@mail.sustech.edu.cn}

\affiliation{$^{a}$Department of Physics, Southern University of Science and Technology, \looseness=-1\\
	Shenzhen 518055, China}

\affiliation{$^{b}$Center for Theoretical Physics, College of Physics, Sichuan University, \looseness=-1\\
	Chengdu 610064, China}

\begin{abstract}
We report on the potential occurrence of a numerical instability in the long-time simulation of black holes using the Baumgarte-Shapiro-Shibata-Nakamura formulation of numerical relativity, even in the simple set-up of a Schwarzschild black hole. Through extensive numerical experiments, we identify that this ``late-time instability'' arises from accumulated violations of the momentum constraint. To address this issue, we propose two modified versions of the so-called conformal covariant Z4 scheme, designed to propagate momentum constraint violations without damping. Our results demonstrate that these alternative formulations, which we refer to as CCZ4' and CCZ3, effectively resolve the late-time numerical instability not only in Schwarzschild spacetimes but also in black hole spacetimes with matter fields. Notably, by preventing damping of the momentum constraint violation, the Hamiltonian constraint damping can be significantly increased, which plays a crucial role in stabilizing long-term evolution in our proposed schemes.
\end{abstract}

\maketitle
\tableofcontents{}

\section{Introduction}

Recent observational breakthroughs in black hole physics \cite{Abbott:2016blz,EventHorizonTelescope:2019dse,EventHorizonTelescope:2022wkp} underscore the need for stable, long-time numerical relativity simulations to study black hole evolution, particularly in systems affected by weak instabilities characterized by very long time scales. Investigating the evolution of such systems through numerical relativity cannot only provide gravitational wave templates for experiments, but also help explore and understand their theoretical properties \cite{Campanelli:2005dd,Pretorius:2005gq,Baker:2005vv,Guo:2024cts}.

A remarkable example of such weak instabilities is the superradiant extraction of mass from a spinning black hole by a massive probe scalar field \cite{Damour:1976kh,Arvanitaki:2009fg}. In this process, the mass of the scalar field serves as a reflective outer boundary, enabling continuous mass extraction from the black hole, leading to the formation of a scalar cloud encircling it \cite{Detweiler:1980uk}. However, the growth rate of such superradiant instabilities is extremely slow, posing significant numerical challenges for studying the long-term evolution of this phenomenon \cite{Cardoso:2004nk}. It has been reported that for a rapidly rotating black hole, the dominant scalar mode associated with this instability exhibits a minimum growth time-scale of order $10^{6}$ in units of the black hole mass \cite{Dolan:2007mj,Dolan:2012yt}. To evade this issue, research on the non-linear evolution of these set-ups has shifted to scenarios involving a charged scalar field around a charged black hole or a massive vector field around a spinning black hole, where the instability rates are significantly faster \cite{Sanchis-Gual:2015lje,East:2017ovw}.\footnote{Besides these technical considerations, charged scalar and vector fields certainly deserve attention due to their physical properties. For instance, hairy black holes formed through the superradiant growth of a charged scalar field are dual to superconductor systems in the context of the AdS/CFT correspondence \cite{Horowitz:2010gk}.} Despite this advantage, the studies still require robust numerical techniques capable of simulating long-term evolution. It has been shown, for instance, that a saturated cloud of massive vectors can radiate slowly through gravitational wave emission over time-scales extending to about $10^{5}M$, as the hole spins down and exits the superradiant regime \cite{East:2017ovw}. In addition to superradiance, weak non-linear instabilities have also been reported in spacetimes featuring a stable light ring \cite{Cardoso:2014sna,Cunha:2022gde}. Numerical studies suggest that exotic compact objects may experience light-ring instabilities after long-term evolution, with time-scales of order $10^{4}M$ \cite{Cunha:2022gde}. Conversely, simulations indicate that black holes with multiple photon spheres appear free of such instabilities, even after extended evolution \cite{Guo:2024cts}, although again diagnosing this outcome is only possible by means of long-term simulations.

One of the most remarkable developments in numerical relativity has been the Baumgarte-Shapiro-Shibata-Nakamura (BSSN) formulation \cite{Shibata:1995we,Baumgarte:1998te}, which is widely used for its accuracy and stability in simulating black hole evolution, particularly for modeling binary mergers and computing gravitational waves \cite{Campanelli:2005dd,Pretorius:2005gq,Baker:2005vv}. The BSSN formalism is based on the construction of a conformal connection as an evolution variable in a specific coordinate system, in order to ensure strong hyperbolicity through the application of the momentum constraint equation. Subsequently, the BSSN formulation has been generalized to a covariant form by introducing a reference metric \cite{Brown:2009dd,Ruchlin:2017com,Baumgarte:2020gwo,Urakawa:2022emf}. This generalized scheme allows for the definition of a covariant and conformal connection, facilitating black hole evolution in coordinate patches that optimize computational efficiency.

On the other hand, the BSSN formulation does not propagate Hamiltonian constraint violations. To address this limitation, the so-called conformal covariant Z4 system (CCZ4), and the related Z4c, have been proposed, in which constraint violations are promoted to dynamical variables \cite{Alic:2011gg,Ruiz:2010qj,Hilditch:2012fp}. The CCZ4 formulation effectively converts the violations of the Hamiltonian and momentum constraints into wave propagation equations. In contrast, the Z4c formulation focuses solely on propagating the Hamiltonian constraint violation, preserving a structure closer to the BSSN formulation \cite{Bernuzzi:2009ex,Hilditch:2012fp}. Similarly to the generalized BSSN formulation, the CCZ4 formulation has also been extended to a fully covariant form by introducing an additional reference metric \cite{Sanchis-Gual:2014nha,Mewes:2020vic,Reid:2023dgm}. While the CCZ4 formulation has demonstrated higher accuracy in simulating binary neutron stars compared to the BSSN formulation,\footnote{See also \cite{Cao:2011fu,Daverio:2018tjf} for detailed comparisons of the performance of BSSN, Z4c and CCZ4.} it suffers from non-linear numerical instabilities when applied to black hole spacetimes. As shown in previous studies \cite{Alic:2013xsa,Sanchis-Gual:2014nha}, these instabilities can be alleviated by adequately introducing damping for the constraint violations. However, a key observation is that the damping effect must not be excessively strong to avoid the emergence of further numerical instabilities (see e.g.\ \cite{Alic:2013xsa} or Fig.\ \ref{Schw_variousScheme} below).

The purpose of this paper is to develop a numerical scheme that ensures stable, long-term evolution and can be applied to studying the ultimate fate of systems afflicted by the aforementioned weak instabilities. The rest of the paper is organized as follows. In Section \ref{secII}, we review several numerical schemes for solving the time evolution of the Einstein equations, and introduce two new proposals specifically tailored for long-term numerical stability, which we dub CCZ4' and CCZ3 schemes, which are characterized by the propagation of the violations of the momentum constraints without damping. In Section \ref{sec:Stable-Evolution-in}, we demonstrate that the CCZ4' and CCZ3 formulations enable stable evolution of a Schwarzschild black hole over time-scales up to $10^{5}$ in units of the black hole's mass. Section \ref{sec:Applications} explores the application of these schemes to the fully non-linear evolution of matter fields in black hole spacetimes, showing that the CCZ3 formulation appears to offer the highest degree of stability and accuracy for studying systems with weak instabilities. We close with a summary of our findings and conclusions in Section \ref{Sec:Conc}. Throughout this paper, we adopt units with $G=c=4\pi\epsilon_{0}=1$.

\section{Numerical schemes for evolving the Einstein equations}
\label{secII}

To numerically evolve the metric variables, we adopt the fully covariant and conformal formulation of the Einstein field equations in the context of the damped Z4 (CCZ4) system \cite{Alic:2011gg,Alic:2013xsa}. The CCZ4 scheme is specifically designed for numerical relativity simulations, where the Einstein equations are reformulated to improve the numerical stability and accuracy. In this system, the Einstein equation is extended to
\begin{equation}
^{(4)}R_{\mu\nu}+2\nabla_{(\mu}{}^{(4)}Z_{\mu)}-\kappa_{1}\left[2n_{(\mu}{}^{(4)}Z_{\nu)}-\left(1+\kappa_{2}\right)g_{\mu\nu}n^{\rho}{}^{(4)}Z_{\rho}\right]=8\pi\left(T_{\mu\nu}-\frac{1}{2}g_{\mu\nu}T\right),\label{eq:Z4-EinsteinEq}
\end{equation}
where $^{(4)}R_{\mu\nu}$ is the four-dimensional
Ricci tensor associated with the metric $g_{\mu\nu}$, $\nabla_{\mu}$ is the covariant derivative, and $T_{\mu\nu}$ is the stress-energy tensor (with $T\equiv T^\mu{}_\mu$). Here, the four-vector $^{(4)}Z_{\mu}$ quantifies deviations from the Einstein equation, particularly important for keeping track of constraint violations during numerical evolution. When $^{(4)}Z_{\mu}=0$, the system reduces to the standard Einstein equation. The constants $\kappa_{1}$ and $\kappa_{2}$ are damping parameters introduced to control the magnitude of $^{(4)}Z_{\mu}$ in numerical simulations. The normal vector to the three-dimensional spatial hypersurface $\Sigma$ is denoted by $n_{\mu}$, pointing along the time evolution of the foliation $\Sigma$ in the future direction. 

In the 3+1 decomposition of spacetime, the line element is expressed in terms of ADM variables as 
\begin{equation}
ds^{2}=-\alpha^{2}dt^{2}+\gamma_{ij}\left(dx^{i}+\beta^{i}dt\right)\left(dx^{j}+\beta^{j}dt\right), \label{metric}
\end{equation}
where $\alpha$ is the lapse function, $\beta^{i}$ is the shift vector, and $\gamma_{ij}$ is the induced physical metric on $\Sigma$. The normal
vector is given by $n_{\mu}=\left(-\alpha,0,0,0\right)$, or $n^{\mu}=\left(1/\alpha,-\beta^{i}/\alpha\right)$. 

Using Bianchi identities, taking the divergence of the modified Einstein equation \eqref{eq:Z4-EinsteinEq} yields a propagation equation for $^{(4)}Z_{\mu}$,
\begin{equation}
\square{}^{(4)}Z_{\mu}=-R_{\mu\nu}{}^{(4)}Z^{\nu}-\kappa_{1}\nabla^{\nu}\left(n_{\mu}{}^{(4)}Z_{\nu}+n_{\nu}{}^{(4)}Z_{\mu}+\kappa_{2}g_{\mu\nu}n_{\sigma}{}^{(4)}Z^{\sigma}\right),\label{eq:Eq4Z}
\end{equation}
which governs the evolution of constraint violations $^{(4)}Z_{\mu}$. In weakly curved spacetimes, this approximates a standard damped wave equation.
In the 3+1 decomposition, $^{(4)}Z_{\mu}$ is split into components along the normal vector $n^{\mu}$ and directions tangential to the spatial hypersurface
$\Sigma$,
\begin{align}
\Theta & =-n_{\mu}{}^{(4)}Z^{\mu},\qquad Z_{i}=\gamma_{i}^{\mu}{}^{(4)}Z_{\mu},\label{eq:Z=000020vec}
\end{align}
where the projection operator is $\gamma_{\mu}^{\nu}=\delta_{\mu}^{\nu}+n_{\mu}n^{\nu}$. Similarly, the stress-energy tensor $T_{\mu\nu}$ is decomposed as
\begin{equation}
\rho=n^{\mu}n^{\nu}T_{\mu\nu},\qquad S_{i}=-\gamma_{i}^{\mu}n^{\nu}T_{\mu\nu},\qquad S_{ij}=\gamma_{i}^{\mu}\gamma_{j}^{\nu}T_{\mu\nu}.
\end{equation}
From Eq.\ \eqref{eq:Eq4Z}, the evolution equations for $\Theta$ and $Z_{i}$ in terms of the ADM variables read
\begin{align}
\partial_{\perp}\Theta & =\frac{\alpha}{2}\left(R+K^{2}-K_{ij}K^{ij}-16\pi\rho\right)-\alpha\Theta K+\alpha D_{i}Z^{i}-Z^{i}\partial_{i}\alpha-\alpha\kappa_{1}\left(2+\kappa_{2}\right)\Theta,\label{eq:eq_Theta}\\
\partial_{\perp}Z_{i} & =\alpha\left(D_{j}K_{\:i}^{j}-D_{i}K-8\pi S_{i}\right)+\alpha\partial_{i}\Theta-2\alpha K_{\:i}^{j}Z_{j}-\Theta\partial_{i}\alpha-\kappa_{1}\alpha Z_{i},\label{eq:eq_Z3}
\end{align}
where $\partial_{\perp}\equiv\partial_{t}-\mathcal{L}_{\beta}$ and $K_{ij}\equiv-\mathcal{L}_{n}\gamma_{ij}/2$ is the extrinsic curvature of the hypersurface $\Sigma$. In the absence of constraint violations (i.e., $^{(4)}Z_{\mu}=0$), the ADM Hamiltonian and momentum constraints of general relativity are recovered as
\begin{align}
H & =\frac{1}{2}\left(R+K^{2}-K_{ij}K^{ij}\right)-8\pi\rho=0,\label{eq:H_cons}\\
M_{i} & =D_{j}K_{\:i}^{j}-D_{i}K-8\pi S_{i}=0.\label{eq:M_cons}
\end{align}
Comparing Eqs.\ \eqref{eq:H_cons} and \eqref{eq:M_cons} with Eqs.\ \eqref{eq:eq_Theta} and \eqref{eq:eq_Z3}, it is evident that $\Theta$ characterizes the Hamiltonian constraint violation while $Z_{i}$ measures the momentum constraint violation. Therefore the evolution equations for $\Theta$ and $Z_{i}$ (Eqs.\ \eqref{eq:eq_Theta} and \eqref{eq:eq_Z3}) describe the propagation of constraint violations during numerical simulations. 

In this paper, we focus on the fully covariant and conformal formulations of numerical relativity. To achieve a conformal formulation, the spatial metric $\gamma_{ij}$ is decomposed as
\begin{equation}
\gamma_{ij}=e^{4\phi}\bar{\gamma}_{ij},
\end{equation}
where $e^{4\phi}$ is the conformal factor. In addition, we impose Brown's Lagrangian condition $\partial_{t}\bar{\gamma}=0$, where $\bar{\gamma}$ represents the determinant of the conformal metric $\bar{\gamma}_{ij}$ \cite{Brown:2009dd}. In this conformal formalism,
the extrinsic curvature is rescaled as 
\begin{equation}
\bar{A}_{ij}=e^{-4\phi}\left(K_{ij}-\frac{1}{3}\gamma_{ij}K\right).
\end{equation}
To implement a fully covariant formulation, a reference metric $\hat{\gamma}_{ij}$ is introduced as a fixed background that remains static throughout the simulation. Using this reference metric, a covariant connection is defined by the difference
\begin{equation}
\Delta\Gamma_{jk}^{i}\equiv\bar{\Gamma}_{jk}^{i}-\hat{\Gamma}_{jk}^{i},
\end{equation}
where $\bar{\Gamma}_{jk}^{i}$ is the connection associated with the conformal metric $\bar{\gamma}_{ij}$ and $\hat{\Gamma}_{jk}^{i}$ is the connection associated with the reference metric $\hat{\gamma}_{ij}$.

Collecting all equations in this formulation we arrive at the following CCZ4 system \cite{Sanchis-Gual:2014nha,Ruchlin:2017com}:
\begin{align}
\partial_{\perp}\bar{\gamma}_{ij}= & \frac{2}{3}\bar{\gamma}_{ij}\left(\alpha\bar{A}_{k}^{k}-\bar{D}_{k}\beta^{k}\right)-2\alpha\bar{A}_{ij},\label{eq:gamma_CCZ4}\\
\partial_{\perp}\bar{A}_{ij}= & e^{-4\phi}\left[-2\alpha\bar{D}_{i}\bar{D}_{j}\phi+4\alpha\bar{D}_{i}\phi\bar{D}_{j}\phi+4\bar{D}_{(i}\alpha\bar{D}_{j)}\phi-\bar{D}_{i}\bar{D}_{j}\alpha+\alpha\left(\bar{R}_{ij}+2D_{(i}Z_{j)}-8\pi S_{ij}\right)\right]^{\textrm{TF}}\nonumber \\
 & -\frac{2}{3}\bar{A}_{ij}\bar{D}_{k}\beta^{k}-2\alpha\bar{A}_{ik}\bar{A}_{\:j}^{k}+\alpha\bar{A}_{ij}\left(K-2\Theta\right),\label{eq:Aij_CCZ4}\\
\partial_{\perp}K= & e^{-4\phi}\left[\alpha\left(\bar{R}-8\bar{D}^{i}\phi\bar{D}_{i}\phi-8\bar{D}^{2}\phi\right)-\left(2\bar{D}^{i}\alpha\bar{D}_{i}\phi+\bar{D}^{2}\alpha\right)\right]+\alpha\left(K^{2}-2\Theta K\right)\nonumber \\
 & +2\alpha D_{i}Z^{i}-3\alpha\kappa_{1}\left(1+\kappa_{2}\right)\Theta+4\pi\alpha\left(S-3\rho\right),\label{eq:K_CCZ4}\\
\partial_{\perp}\phi= & \frac{1}{6}\bar{D}_{k}\beta^{k}-\frac{1}{6}\alpha K,\label{eq:phi_CCZ4}\\
\partial_{\perp}\Theta= & \frac{1}{2}\alpha\left[e^{-4\phi}\left(\bar{R}-8\bar{D}^{i}\phi\bar{D}_{i}\phi-8\bar{D}^{2}\phi\right)-\bar{A}_{ij}\bar{A}^{ij}+\frac{2}{3}K^{2}-2\Theta K+2D_{i}Z^{i}\right],\label{eq:Theta_CCZ4}\\
\partial_{\perp}\tilde{\Lambda}^{i}= & \bar{\gamma}^{jk}\hat{D}_{j}\hat{D}_{k}\beta^{i}+\frac{2}{3}\tilde{\Lambda}^{i}\bar{D}_{j}\beta^{j}+\frac{1}{3}\bar{D}^{i}\bar{D}_{j}\beta^{j}-2\bar{A}^{ij}\left(\partial_{j}\alpha-6\alpha\partial_{j}\phi\right)+2\alpha\bar{A}^{jk}\Delta\Gamma_{jk}^{i}\nonumber \\
 & -\frac{4}{3}\alpha\bar{\gamma}^{ij}\partial_{j}K+2\bar{\gamma}^{ij}\left(\alpha\partial_{j}\Theta-\Theta\partial_{j}\alpha-\frac{2}{3}\alpha KZ_{j}\right)-2\alpha\kappa_{1}\bar{\gamma}^{ij}Z_{j}-16\pi\alpha\bar{\gamma}^{ij}S_{j},\label{eq:Lambda_CCZ4}
\end{align}
where `$\textrm{TF}$' denotes the trace-free part of the corresponding tensor. Here, $\hat{D}_{i}$, $\bar{D}_{i}$ and $D_{i}$ represent the covariant derivatives with respect to the reference metric $\hat{\gamma}_{ij}$, the conformal metric $\bar{\gamma}_{ij}$, and the ADM metric $\gamma_{ij}$, respectively. In Eq.\ \eqref{eq:Lambda_CCZ4}, the dynamical variable
$\tilde{\Lambda}^{i}$ is defined as
\begin{equation}
\tilde{\Lambda}^{i}\equiv\bar{\Lambda}^{i}+2\bar{\gamma}^{ij}Z_{j},
\end{equation}
with $\bar{\Lambda}^{i}\equiv\bar{\gamma}^{jk}\Delta\Gamma_{jk}^{i}$. We remark that the evolution equation for $\bar{\Lambda}^{i}$ can also be derived from the momentum constraint equation, indicating it is not independent of Eq.\ \eqref{eq:eq_Z3}. In the CCZ4 formulation, Eq.\ \eqref{eq:Lambda_CCZ4} governing $\tilde{\Lambda}^{i}$ serves two purposes: it functions like $\bar{\Lambda}^{i}$ in the BSSN formulation for constructing a strongly hyperbolic system and also propagates momentum constraint violations as in Eq.\ \eqref{eq:eq_Z3}.

Regarding the fixing of gauge, we employ the $1+\textrm{log}$ slicing condition for the lapse field $\alpha$ and the $\Gamma$-driver condition for the shift vector field $\beta^{i}$ \cite{Bona:1994dr,Alcubierre:2002kk,Schnetter:2010cz,Ruchlin:2017com},
\begin{align}
\partial_{\perp}\alpha & =-2\alpha\left(K-2\Theta\right),\\
\partial_{\perp}\beta^{i} & =C^{i},\nonumber \\
\partial_{\perp}C^{i} & =\frac{3}{4}\partial_{\perp}\tilde{\Lambda}^{i}-\eta C^{i}, \label{eq:gamma driver}
\end{align}
where $C^i$ is an auxiliary field and $\eta$ is another damping parameter, and for simplicity we choose $\eta=1$ throughout the main text of this paper. Modifications to the form of the $\Gamma$-driver condition are discussed briefly in Appendix \ref{app:gamma}.

In accordance with our aim of comparing different numerical schemes, we introduce the following schemes as alternatives to the CCZ4 formulation outlined above:
\begin{lyxlist}{00.00.0000}
\item [{\textbf{BSSN}:}] This formulation is composed of Eqs.\ \eqref{eq:gamma_CCZ4},
\eqref{eq:Aij_CCZ4}, \eqref{eq:phi_CCZ4},
and together with 
\begin{align}
\partial_{\perp}K &= \frac{1}{3}\alpha K^{2}+\alpha\bar{A}_{ij}\bar{A}^{ij}-e^{-4\phi}\left(2\bar{D}^{i}\alpha\bar{D}_{i}\phi+\bar{D}^{2}\alpha\right),\label{eq:K_BSSN}\\
\partial_{\perp}\bar{\Lambda}^{i} &= \bar{\gamma}^{jk}\hat{D}_{j}\hat{D}_{k}\beta^{i}+\frac{2}{3}\bar{\Lambda}^{i}\bar{D}_{j}\beta^{j}+\frac{1}{3}\bar{D}^{i}\bar{D}_{j}\beta^{j}-2\bar{A}^{ij}\left(\partial_{j}\alpha-6\alpha\partial_{j}\phi\right)\nonumber \\
 & \quad+2\alpha\bar{A}^{jk}\Delta\Gamma_{jk}^{i}-\frac{4}{3}\alpha\bar{\gamma}^{ij}\partial_{j}K-16\pi\alpha\bar{\gamma}^{ij}S_{j}.\label{eq:Lambda_BSSN}
\end{align}
In this formulation, $\Theta$ and $Z_{i}$ are set to zero in all equations.

\item [{\textbf{CCZ4'}:}] This formulation follows the equations of the CCZ4 systems as defined in Eqs.\ \eqref{eq:gamma_CCZ4}, \eqref{eq:Aij_CCZ4}, \eqref{eq:K_CCZ4}, \eqref{eq:phi_CCZ4}, \eqref{eq:Theta_CCZ4} and \eqref{eq:Lambda_CCZ4}. However, the damping factor $\kappa_{1}$ is replaced by $\kappa_{\Theta}$ in Eq.\ \eqref{eq:K_CCZ4} and by $\kappa_{\Gamma}$ in Eq.\ \eqref{eq:Lambda_CCZ4}. In other words, we allow for different damping factors in these two equations.

\item [{\textbf{CCZ0}:}] This formulation consists of the same equations as the CCZ4 system: \eqref{eq:gamma_CCZ4}, \eqref{eq:Aij_CCZ4}, \eqref{eq:K_CCZ4}, \eqref{eq:phi_CCZ4}, \eqref{eq:Theta_CCZ4} and \eqref{eq:Lambda_CCZ4}, except that $Z_{i}=0$ is set in all equations.

\item [{\textbf{CCZ3}:}] This formulation uses the evolution equations
\eqref{eq:gamma_CCZ4}, \eqref{eq:Aij_CCZ4}, \eqref{eq:K_CCZ4}, \eqref{eq:phi_CCZ4}
and \eqref{eq:Lambda_CCZ4}. In this formulation, $\Theta=0$ is set  in all equations.
\end{lyxlist}

The CCZ4' and CCZ3 formulations correspond to our new proposals, which we will show to be effective for the purpose of long-term black hole evolution and the related question of numerical stability. On the other hand, the BSSN and CCZ0 will instead be used for the purpose of comparison and for explaining the advantages of our formulations.

\section{Stable evolution of black hole spacetimes}
\label{sec:Stable-Evolution-in}

In this Section, we study the application of the numerical schemes discussed in the previous Section to long-term simulations of black hole evolution. Numerical experiments with a Schwarzschild black hole reveal that the standard numerical scheme, specifically the BSSN formulation, exhibits numerical instabilities during long-time simulations. We trace the source of these instabilities to violations
of the momentum constraint. We then show that our modified numerical schemes, namely CCZ4' and CCZ3, succeed in addressing this issue by ensuring long-term stability in black hole simulations.

We integrate the numerical framework reviewed in Section \ref{secII} into the \textit{BlackHoles@Home} platform \cite{BHathome}. For practical implementation, we employ a reference metric given by
\begin{equation}
\hat{dl}^{2}=dr^{2}+r^{2}d\Omega=(dr/dR)^{2}dR^{2}+r(R)^{2}d\Omega,
\end{equation}
in spherical-like coordinates $(R,\theta,\varphi)$. Here, the radial
coordinate $r$ is scaled to a dimensionless quantity $R$ by
\begin{equation}
r=r_{max}\left(R R_{0}+\frac{e^{R/a}-e^{-R/a}}{e^{1/a}-e^{-1/a}}\right),\label{eq:R_r}
\end{equation}
where $r_{max}$ denotes the outer boundary in the numerical simulation, and $R_{0}$ and $a$ are scaling factors linking $r$ and $R$. This transformation maps the radial range $[0,r_{max}(R_0 +1)]$ onto $[0,1]$, and note that $r_{max}(R_0 +1)\approx r_{max}$ since $R_0$ is chosen to be very small. Notice that we restrict our attention to spherically symmetric systems in this paper. Radial discretization is achieved using a uniform grid in the rescaled coordinate $R$, with a grid size of $N_{R}$ cells. To stabilize the simulations and mitigate high-frequency numerical noise, we apply the Kreiss-Oliger (KO) dissipation technique to all evolved variables. Additionally, we consider the Courant-Friedrichs-Lewy (CFL) condition (see e.g.\ \cite{Press:2007ipz}), defined by the factor 
\begin{equation}
{\rm CFL}=\frac{\Delta t}{\Delta r_{min}},
\end{equation}
where $\Delta r_{min}$ is the smallest spatial step in the grid. To maintain numerical stability, the CFL factor must not exceed one, ensuring that information does not propagate more than one grid cell during a single time step. We use fourth-order finite differential on the spatial direction, and forth-order Runge-Kutta for the time integration. 

The benchmark parameters used in our simulations are as follows:
\begin{equation}
r_{max}=60000M,\;R_{0}=0.00012,\;a=0.07,\;N_{R}=300,\;\epsilon_{\rm KO}=0.2,\; {\rm CFL}=1.0,\label{eq:setting}
\end{equation}
where $\epsilon_{\rm KO}$ specifies the strength of the KO dissipation. Here and in what follows, $M$ is the initial mass of the black hole. Unless otherwise stated, these parameters \eqref{eq:setting} are used as the default settings for numerical evolution in this study. In the following, we simulate the area of the apparent horizon $A_{h}$ for the purpose of examining the numerical stability.

\subsection{Late-time numerical instability}

\begin{figure}[t]
\begin{centering}
\includegraphics[scale=0.8]{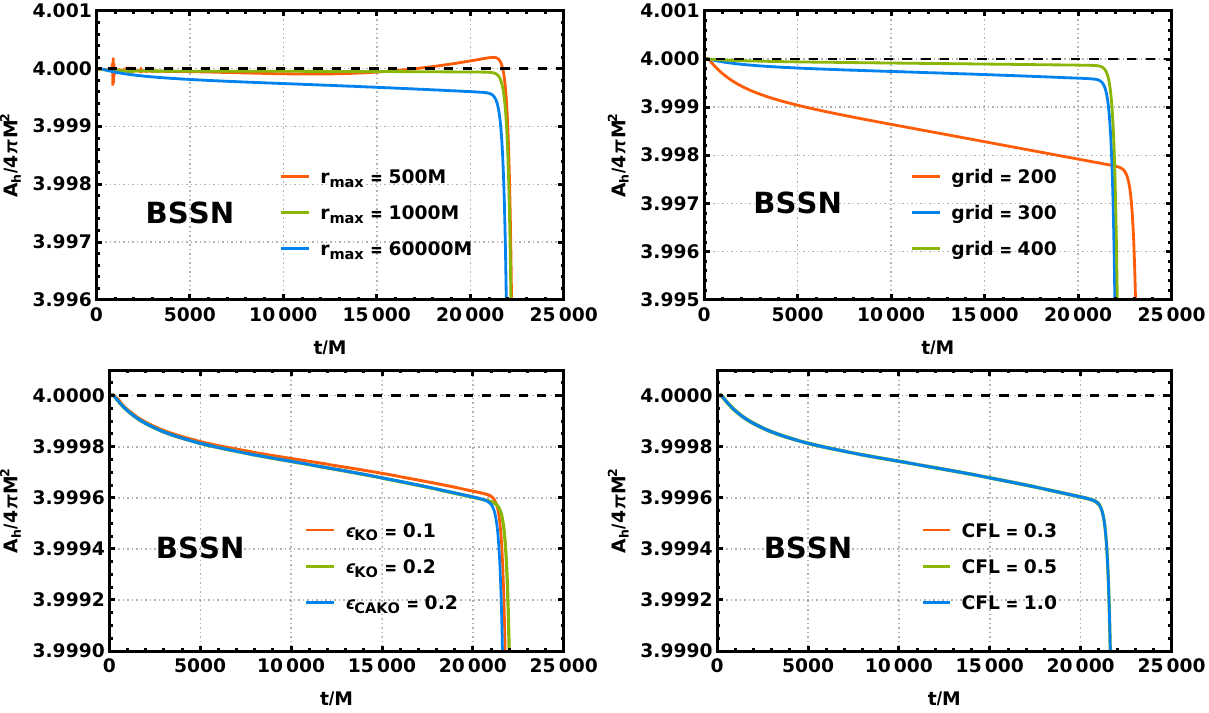}
\par\end{centering}
\caption{Schwarzschild black hole evolution, as measured by the area of the apparent horizon, $A_h$, under various parameter settings. These include different scaling parameters (upper-left panel), varying grid numbers (upper-right), applying different dissipation schemes (lower-left), and different CFL factors (lower-right). See the main text for further details. The results demonstrate that the numerical instability persists at late times irrespective of parameter adjustments.}
\label{Schw_BSSNs}
\end{figure}

For the numerical evolution of an isolated black hole, we employ pre-collapsed initial data representing a Schwarzschild black hole. This configuration transitions from a wormhole slice to a trumpet slice using the moving puncture method \cite{Campanelli:2005dd,Baker:2005vv,Brown:2009ki}. The initial data for the spacetime is given by 
\begin{equation}
e^{\phi}=1+\frac{M}{2r},
\end{equation}
with pre-collapsed lapse $\alpha=e^{-2\phi}$ and vanishing shift $\beta^{i}=0$.

In Fig.\ \ref{Schw_BSSNs}, we present the results of our simulations of the evolution of a Schwarzschild black hole within the BSSN formulation, using various numerical parameter settings. The upper-left panel compares different scaling parameters, where $r_{max}=500M,\;R_{0}=0.0015,\;a=0.15$ (orange line); $r_{max}=1000M,\;R_{0}=0.001,\;a=0.12$ (green); and $r_{max}=60000M,\;R_{0}=0.0012,\;a=0.07$ (blue). For smaller values of $r_{max}$, noticeable numerical perturbations emerge due to noise from the outer boundary during the simulation. The upper-right panel varies grid numbers: $N_{R}=200$ (orange), $N_{R}=300$ (blue) and $N_{R}=400$ (green). Simulations with fewer grid points ($N_{R}=200$) exhibit more pronounced numerical errors compared to higher-resolution set-ups ($N_{R}=300$ or $400$). The lower-left panel examines dissipation schemes with $\epsilon_{\rm KO}=0.1$ (orange), $\epsilon_{\rm KO}=0.2$ (green) and $\epsilon_{\rm CAKO}=0.2$ (blue). Here, $\epsilon_{\rm CAKO}=e^{-2\phi}\epsilon_{\rm KO}$ is the curvature-adjusted KO dissipation strength, which was proposed to reduce numerical noise near the puncture by an additional dissipation term \cite{Etienne:2024ncu}. Lastly, the dependence on the CFL factor is tested in the lower-right panel: $\rm CFL=0.3$ (orange), $\rm CFL=0.5$ (green), $\rm CFL=1.0$ (blue); the variation in this case is essentially negligible, as seen in the graph where all three curves overlap.

The clear lesson revealed by the simulations is the presence of a late-time numerical instability that persists for a broad range of parameter adjustments. These findings suggest that this ``late-time instability'' is an essential aspect of the numerical formulation rather than the parameter settings.

\subsection{Modified numerical schemes}

\begin{figure}[t]
\begin{centering}
\includegraphics[scale=0.8]{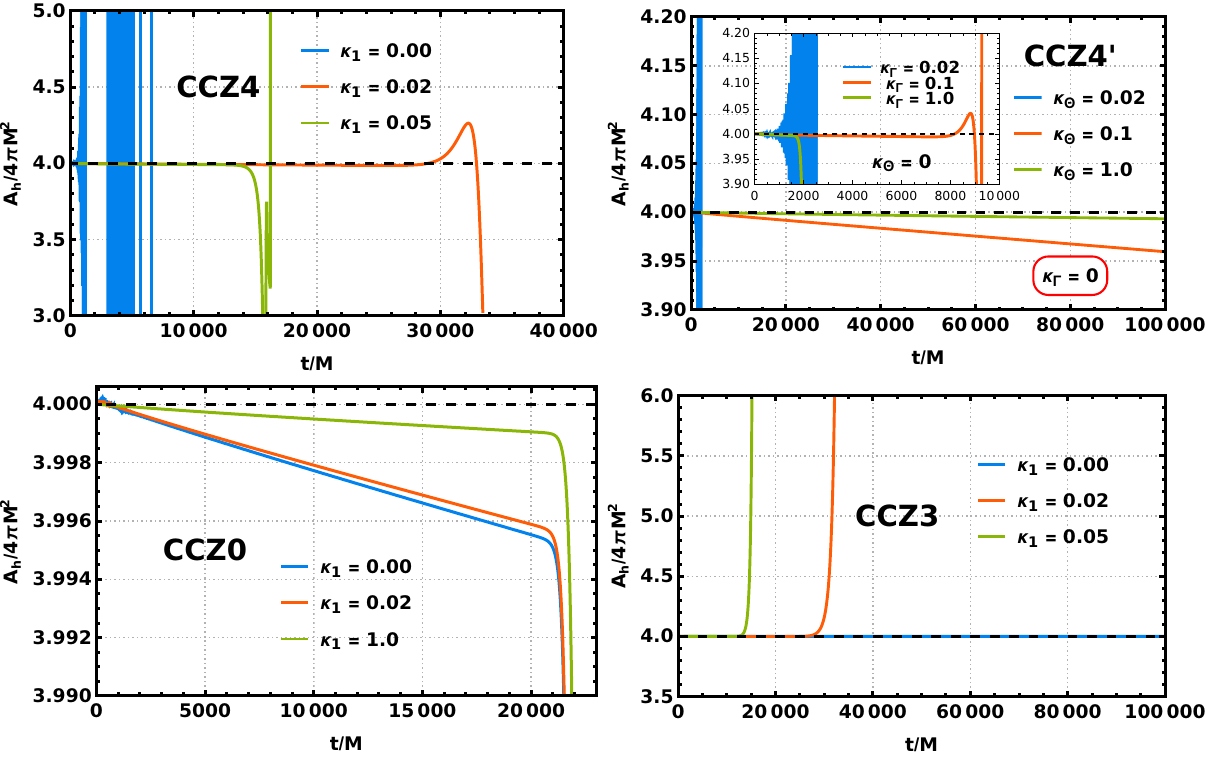}
\par\end{centering}
\caption{Numerical evolution of a Schwarzschild black hole, as measured by the area of the apparent horizon, $A_h$, under different numerical schemes. {\it Upper-left panel:} The CCZ4 formulation demonstrates that $\kappa_{1}$ impacts numerical stability, with nonlinear instabilities suppressed by moderate damping but exacerbated by excessive damping. {\it Upper-right:} The modified CCZ4' formulation considers two different damping parameters in place of $\kappa_{1}$, i.e.\ $\kappa_{\Theta}$ and $\kappa_{\Gamma}$, with the results showing that suppressing momentum constraint violations ($\kappa_{\Gamma}=0$) stabilizes the simulation, while omitting Hamiltonian constraint damping causes instabilities. {\it Lower-left:} The CCZ0 formulation, which disables momentum constraint propagation ($Z_{i}=0$), is seen to exhibit a ``late-time instability'' similar to the BSSN formulation. {\it Lower-right:} The CCZ3 formulation, defined by the disabling of the propagation of the Hamiltonian constraint ($\Theta=0$), effectively removes late-time instabilities in the absence of damping, but is seen to introduce new instabilities when damping terms are applied.}
\label{Schw_variousScheme}
\end{figure}

In the BSSN formulation, constraint violations are not part of the evolution equations of the system. Numerical errors, which are unavoidable in practical computations, can arise from truncation in finite differencing during spatial discretization, inaccuracies in time integration using the Runge-Kutta algorithm, or precision loss after numerous iterations. Without appropriate mitigation, these errors can accumulate, specifically during black hole evolution, causing constraint violations ($\Theta$ and $Z_{i}$) to grow significantly, potentially leading to numerical instabilities at late times. Thus we anticipate that the ``late-time instability'' we have identified already in the simple set-up of a Schwarzschild black hole, within the BSSN formulation, may potentially be removed by employing the CCZ4 formulation, or its modified versions proposed here, which are designed to propagate constraint violations.
\begin{figure}[t]
\begin{centering}
\includegraphics[scale=0.8]{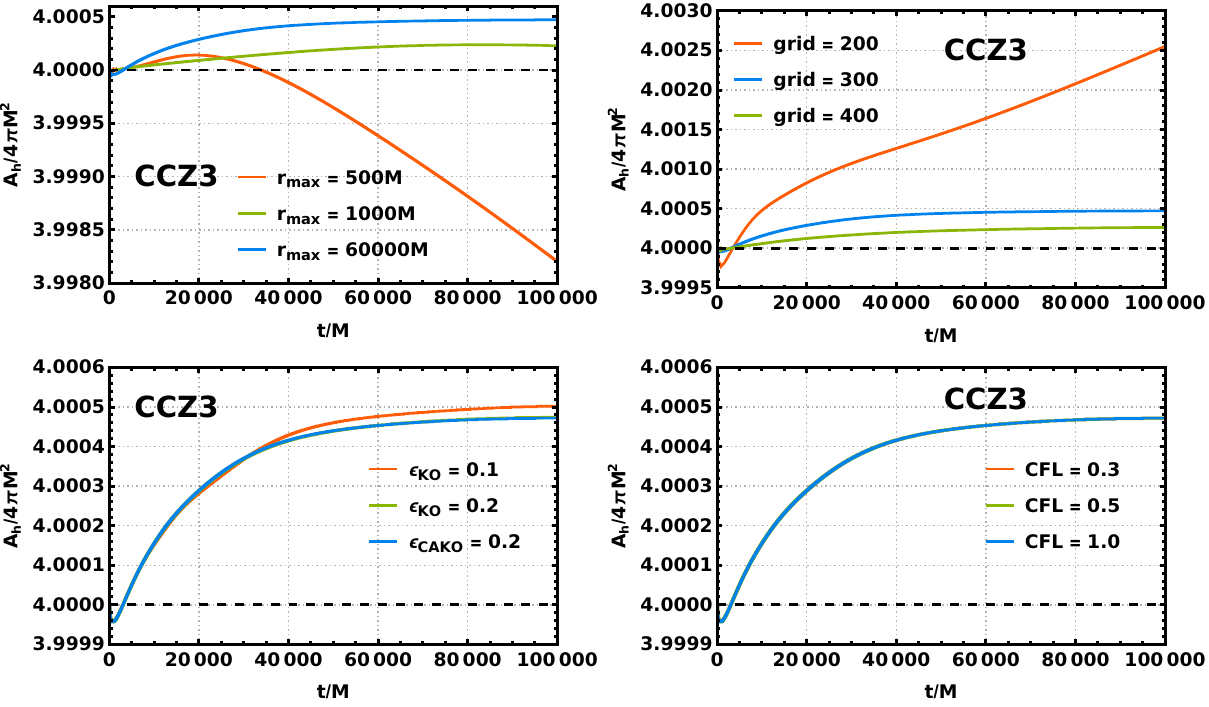}
\par\end{centering}
\caption{Numerical evolution of a Schwarzschild black hole, as measured by the area of the apparent horizon, $A_h$, using the CCZ3 formulation. The ``late-time instability'' is seen to be absent for all parameter settings used in the previous analysis, cf.\ Fig.\ \ref{Schw_BSSNs}, with numerical errors remaining small throughout the simulation.}
\label{Schw_CCZ3}
\end{figure}

In Fig.\ \ref{Schw_variousScheme}, we present the numerical evolution of a Schwarzschild black hole using different numerical schemes, where in all cases we have set $\kappa_{2}=0$ (a choice which ensures appropriate damping \cite{Gundlach:2005eh} and is commonly made in the literature). In the upper-left panel, the results indicate that $\kappa_{1}$ has a significant impact on the numerical stability in the CCZ4 formulation. Non-linear effects arising from constraint violations can destabilize the simulation, as shown by the blue curve. As reported in \cite{Alic:2013xsa,Sanchis-Gual:2014nha}, introducing the damping factor $\kappa_{1}$ indeed helps alleviate these instabilities. However, excessively large values of $\kappa_{1}$ can lead to other numerical instabilities, as illustrated by the green line ($\kappa_{1}=0.05$). While the CCZ4 formulation offers slight improvements in long-term stability over the BSSN formulation, it is clear that it does not fully resolve the ``late-time  instability'' issue.

The upper-right panel examines our first modified scheme, what we refer to as CCZ4' formulation, where $\kappa_{1}$ is replaced by $\kappa_{\Theta}$, which damps the Hamiltonian constraint violation $\Theta$, and by $\kappa_{\Gamma}$, which damps momentum constraint violations $Z_{i}$. Without damping of the momentum constraint violation ($\kappa_{\Gamma}=0$), the numerical instability is found to be suppressed for sufficiently large values of $\kappa_{\Theta}$. On the other hand, as shown in the inset, excluding the damping effect for Hamiltonian constraint violations results in unstable simulations regardless of the choice of $\kappa_{\Gamma}$. This result suggests that, in the CCZ4' formulation, the absence of damping for momentum constraint violations allows for a large $\kappa_{\Theta}$ to be effective in removing late-time instabilities.

The lower-left panel presents simulations using the CCZ0 formulation, where momentum constraint propagation is disabled by setting $Z_{i}=0$. The numerical behavior closely resembles that of the BSSN formulation in Fig.\ \ref{Schw_BSSNs}, with the presence of late-time instabilities. This outcome clearly shows that the evolution of the breaking of the momentum constraint is key to the issue of long-term evolution.

Finally, in the lower-right panel, we study the scheme that we refer to as CCZ3, where the Hamiltonian
constraint propagation is disabled by setting $\Theta=0$. The numerical results show that the ``late-time instability'' can be effectively eliminated when the damping term is absent ($\kappa_{1}=0$). However, introducing a damping term leads to the reappearance of numerical instabilities at late times.

To further investigate the robustness of the CCZ3 formulation, we present in Fig.\ \ref{Schw_CCZ3} numerical simulations  of a Schwarzschild black hole, employing the same parameter settings as in Fig.\ \ref{Schw_BSSNs}. The results make it manifest that the ``late-time instability'' is absent in all cases, with numerical errors remaining under control and within a narrow range of deviation. Together with the outcomes of the above analysis, we can infer that the late-time numerical destabilization observed in the previous Section may be attributed to numerical noise arising from violations of the momentum constraint. It is noteworthy that applying a damping effect to the momentum constraint deviation can induce non-linear numerical instabilities, even with small damping values. The clear conclusion is that the evolution of the momentum constraint violations $Z_{i}$, without the introduction of damping terms, is key for achieving robust long-term simulations of black holes.

\section{Applications to non-vacuum black holes}
\label{sec:Applications}

We study next the application of the numerical schemes discussed in Section \ref{sec:Stable-Evolution-in} to the evolution of black hole spacetimes in the presence of matter fields, specifically focusing on the Reissner-Nordstr\"om (RN) black hole and a black hole that undergoes spontaneous scalarization within the Einstein-Maxwell-scalar (EMS) system \cite{Herdeiro:2018wub,Guo:2024cts,Garcia-Saenz:2024beb}. Our numerical results strongly suggest that the CCZ3 formulation is the most robust and accurate for simulating long-term black hole evolution in the presence of matter fields.

\subsection{Numerical evolution of matter fields}

In the EMS system, a scalar field $\Phi$ (here assumed real for simplicity) is minimally coupled to gravity and non-minimally coupled to the electromagnetic field $A_{\mu}$,
\begin{equation}
S=\frac{1}{16\pi}\int d^{4}x\sqrt{-g}\left[R-2\partial_{\mu}\Phi\partial^{\mu}\Phi-f\left(\Phi\right)F^{\mu\nu}F_{\mu\nu}\right],\label{eq:action}
\end{equation}
where $F_{\mu\nu}=\partial_{\mu}A_{\nu}-\partial_{\nu}A_{\mu}$. The coupling function is chosen as $f\left(\Phi\right)=e^{\alpha_{0}\Phi^{2}}$, a choice that ensures the existence of tachyonic-type instabilities in the vicinity of a RN black hole, provided the coupling constant $\alpha_0$ is large enough \cite{Herdeiro:2018wub}. Varying the action \eqref{eq:action} with respect to the scalar and electromagnetic fields yields their equations of motion,
\begin{align}
\square\Phi & =\frac{1}{4}\dot{f}\left(\Phi\right)F^{2},\label{eq:scalarEq}\\
\nabla_{\mu}F^{\mu\nu} & =-\partial_{\mu}f\left(\Phi\right)F^{\mu\nu},\label{eq:EMfield}
\end{align}
where $\dot{f}\left(\Phi\right)=df\left(\Phi\right)/d\Phi$. The corresponding stress-energy tensor sourcing the Einstein equation \eqref{eq:Z4-EinsteinEq} is given by
\begin{equation}
T_{\mu\nu}=\frac{1}{8\pi}\left[2\partial_{\mu}\Phi\partial_{\nu}\Phi-g_{\mu\nu}\partial_{\rho}\Phi\partial^{\rho}\Phi+f\left(\Phi\right)\left(2F_{\mu\rho}F_{\nu}^{\:\rho}-\frac{1}{2}g_{\mu\nu}F^{2}\right)\right].
\end{equation}

The electromagnetic field also obeys a constraint equation given by the Gauss law: $\nabla_{[\rho}F_{\mu\nu]}=0$. Similar to the CCZ4 approach to the Einstein equations, one can introduce quantities measuring the violation of the constraints, respectively $\Psi_{E}$ for the electric field and $\Psi_{B}$ for the magnetic field, by extending the electromagnetic field equations as \cite{Hirschmann:2017psw}
\begin{align}
\nabla^{\mu}\left(F_{\mu\nu}+g_{\mu\nu}\Psi_{E}\right) & =-\partial_{\mu}f\left(\Phi\right)F^{\mu\nu}+\kappa_{E}n_{\nu}\Psi_{E},\nonumber \\
\nabla^{\mu}\left(*F_{\mu\nu}+g_{\mu\nu}\Psi_{B}\right) & =\kappa_{B}n_{\nu}\Psi_{B},\label{eq:EM_Eqs}
\end{align}
where $*F_{\mu\nu}$ is the Hodge dual of $F_{\mu\nu}$, and $\kappa_{E}$ and $\kappa_{B}$ are damping factors aimed at reducing the amount of constraint violations. 

Analogously to the projections defined in Eq.\ \eqref{eq:Z=000020vec}, we decompose the Maxwell field into electric and magnetic fields via
\begin{equation}
E_{i}=\gamma_{i}^{\mu}n^{\nu}F_{\mu\nu},\qquad B_{i}=\gamma_{i}^{\mu}n^{\nu}\left(*F_{\mu\nu}\right).
\end{equation}
In relation to the scalar field, we also introduce the variable $\Pi=n^{\mu}\nabla_{\mu}\Phi$, which acts as the momentum of the field. The collected evolution equations for the matter fields are then given by
\begin{align}
\partial_{\perp}\Phi & =\alpha\Pi,\nonumber \\
\partial_{\perp}\Pi & =D^{i}\left(\alpha D_{i}\Phi\right)+\alpha\Pi K-\frac{\alpha}{4}\dot{f}\left(\Phi\right)F^{2},\nonumber \\
\partial_{\perp}E^{i} & =\alpha KE^{i}-\alpha D^{i}\Psi_{E}+\epsilon^{ijk}D_{j}\left(\alpha B_{k}\right)+\alpha\frac{\dot{f}\left(\Phi\right)}{f\left(\Phi\right)}\left(\epsilon^{ijk}\partial_{j}\Phi B_{k}-E^{i}\Pi\right),\nonumber \\
\partial_{\perp}\Psi_{E} & =-\alpha\left(\frac{\dot{f}}{f}D_{i}\phi E^{i}+D_{i}E^{i}+\kappa_{E}\Psi_{E}\right),\nonumber \\
\partial_{\perp}B^{j} & =-\epsilon^{jkl}D_{k}\left(\alpha E_{l}\right)+\alpha\left(KB^{j}+D^{j}\Psi_{B}\right),\nonumber \\
\partial_{\perp}\Psi_{B} & =\alpha\left(D_{i}B^{i}-\kappa_{B}\Psi_{B}\right).
\end{align}

\subsection{RN black hole}

\begin{figure}[t]
\begin{centering}
\includegraphics[scale=0.8]{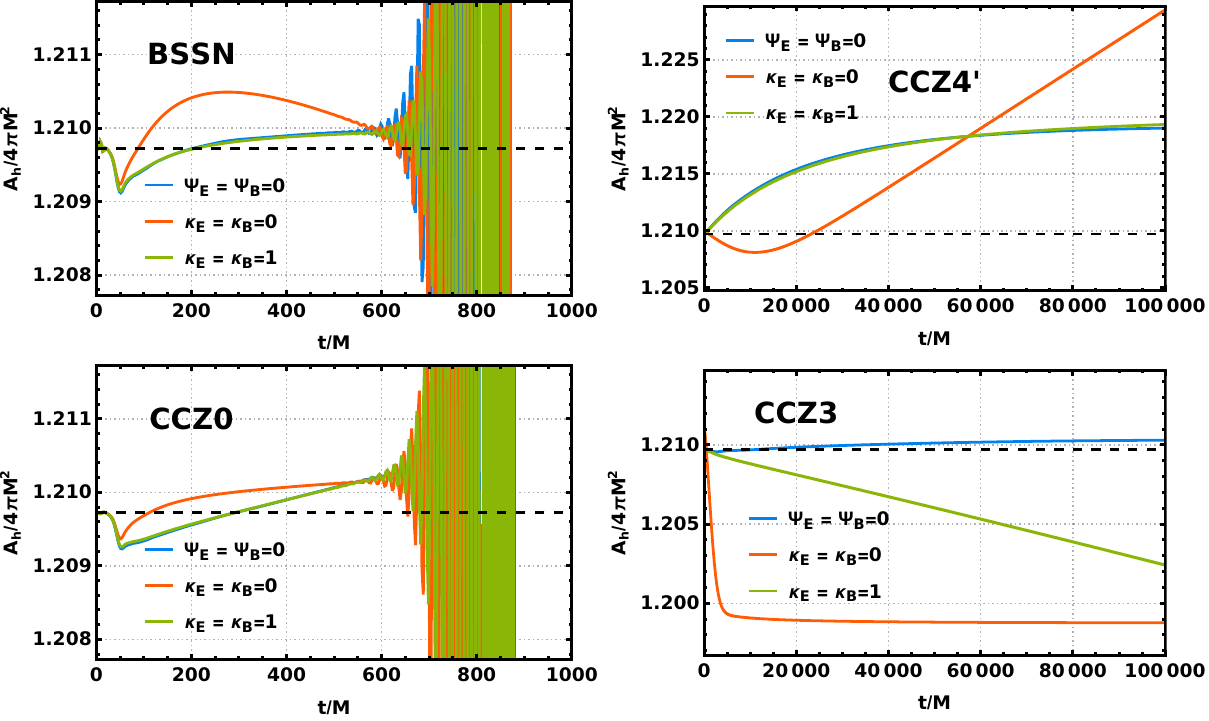}
\par\end{centering}
\caption{Evolution of a RN black hole, as measured by the area of the apparent horizon, $A_h$, with $Q/M=0.995$ and using various numerical schemes. The simulations demonstrate that the BSSN (upper-left panel) and CCZ0 (lower-left) formulations exhibit numerical instabilities at late times, while the CCZ4' (upper-right) and CCZ3 (lower-right) formulations propagate the constraint violations in a stable manner, maintaining accuracy without late-time divergences.}
\label{RN_BH}
\end{figure}

Focusing first on the electric RN black hole spacetime, the initial data is given by \cite{Alcubierre:2009ij}
\begin{equation}
e^{2\phi}=\left(1+\frac{M}{2r}\right)^2-\frac{Q^{2}}{4r^{2}},\quad E^{r}=e^{-6\phi}\frac{Q}{r^{2}},\quad B^{i}=0,\quad\Phi=0.\label{eq:ID_RN}
\end{equation}
In Fig.\ \ref{RN_BH} we present the results of several simulations of this spacetime, considering a large charge-to-mass ratio ($Q/M=0.995$) in order to appreciate the differences relative to the vacuum case studied in Section \ref{sec:Stable-Evolution-in}. The four panels in the figure correspond to the different numerical schemes: BSSN (upper-left panel), CCZ4' (upper-right), CCZ0 (lower-left) and CCZ3 (lower-right). Similarly to the simulation of a Schwarzschild black hole, the BSSN and CCZ0 formulations fail to maintain a stable numerical evolution at late times, exhibiting oscillating divergences. In contrast, the CCZ4' and CCZ3 formulations successfully simulate the evolution without encountering late-time instabilities, once again highlighting the virtue of the propagation of the momentum constraint $Z_{i}$ in long-term simulations.

It is noteworthy that the propagation of constraint violations in the electromagnetic system does not lead to significant improvement in numerical stability. In particular, the CCZ3 formulation is observed to yield the most accurate results in the case where the electromagnetic constraints are defined as non-propagating, i.e.\ $\Psi_{E}=\Psi_{B}=0$. This appears to apply in general, as suggested by similar findings in the hairy black hole model studied in the next subsection.

\subsection{Spontaneous scalarization}

\begin{figure}[t]
\begin{centering}
\includegraphics[scale=0.8]{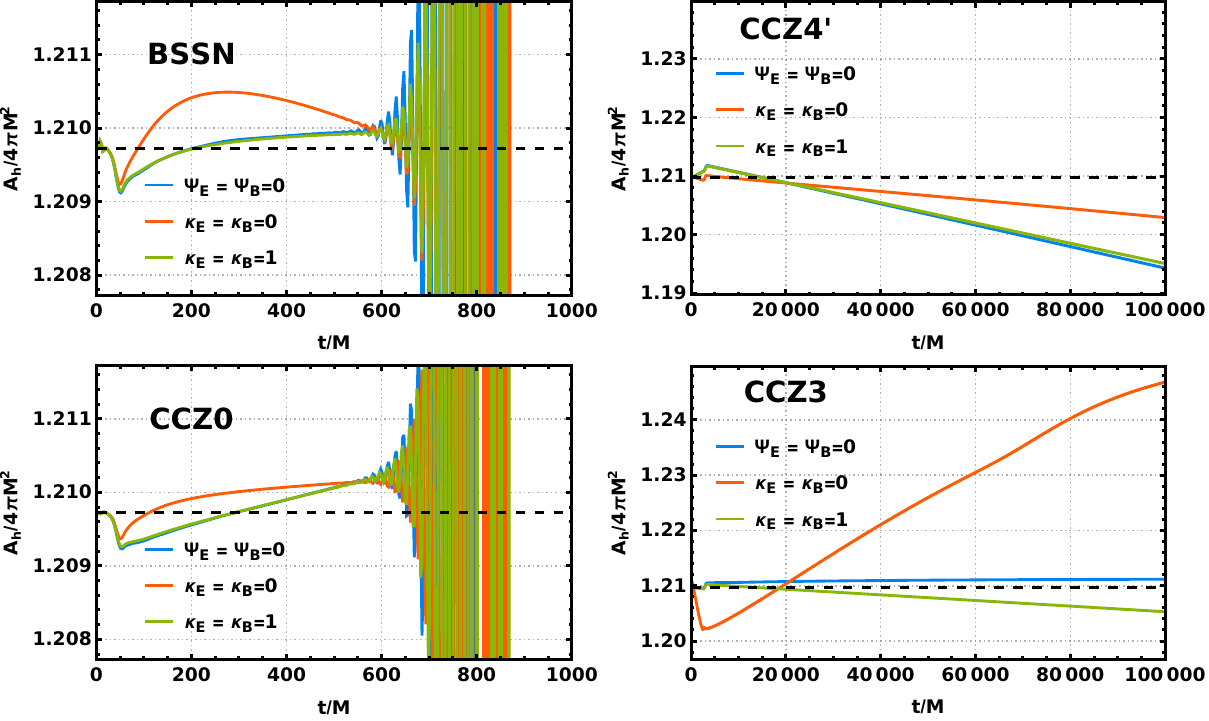}
\par\end{centering}
\caption{The area of the apparent horizon, $A_{h}$, during the evolution of spontaneous scalarization in the EMS model with coupling constant $\alpha_0=1$. The evolution starts from the RN black hole with $Q/M=0.995$ and an initial perturbation of the scalar field given by a Gaussian wave-packet (see the main text). The results indicate that only the CCZ4' (upper-right panel) and CCZ3 (lower-right) formulations yield stable long-term evolution and correctly simulate the formation of a scalarized black hole, whereas the BSSN (upper-left) and CCZ0 (lower-left) schemes fail to accurately do so.}
\label{SBH_AH}
\end{figure}

We next simulate the evolution of the spontaneous scalarization mechanism of a charged black hole in the EMS model. The simulations start with a RN black hole background with the initial data of Eq.\ \eqref{eq:ID_RN}, along with a small initial perturbation of the scalar field corresponding to a spherical Gaussian wave-packet: $\delta\Phi=pe^{-\frac{r^{2}}{M^2}}$, with $p=10^{-4}$. The coupling constant is set to $\alpha_0=1$, ensuring the scalarization rate is slow enough to call for robust long-term numerics. Fig.\ \ref{SBH_AH} displays the numerical evolution in the time domain of the system, in which the initial RN black hole evolves into a scalarized state with scalar hair. The numerical schemes are the same as in Fig.\ \ref{RN_BH}. Similarly to the vacuum and electro-vacuum cases, our results show that the BSSN and CCZ0 formulations do not provide robust results in the long term, failing in this case to fully capture the formation of the scalarized black hole, with the evolution diverging in a comparatively short time. In contrast, both the CCZ4' and CCZ3 formulations successfully establish stable long-term evolutions for the spontaneous scalarization process triggered by weak physical instabilities.

\begin{figure}[t]
\begin{centering}
\includegraphics[scale=0.8]{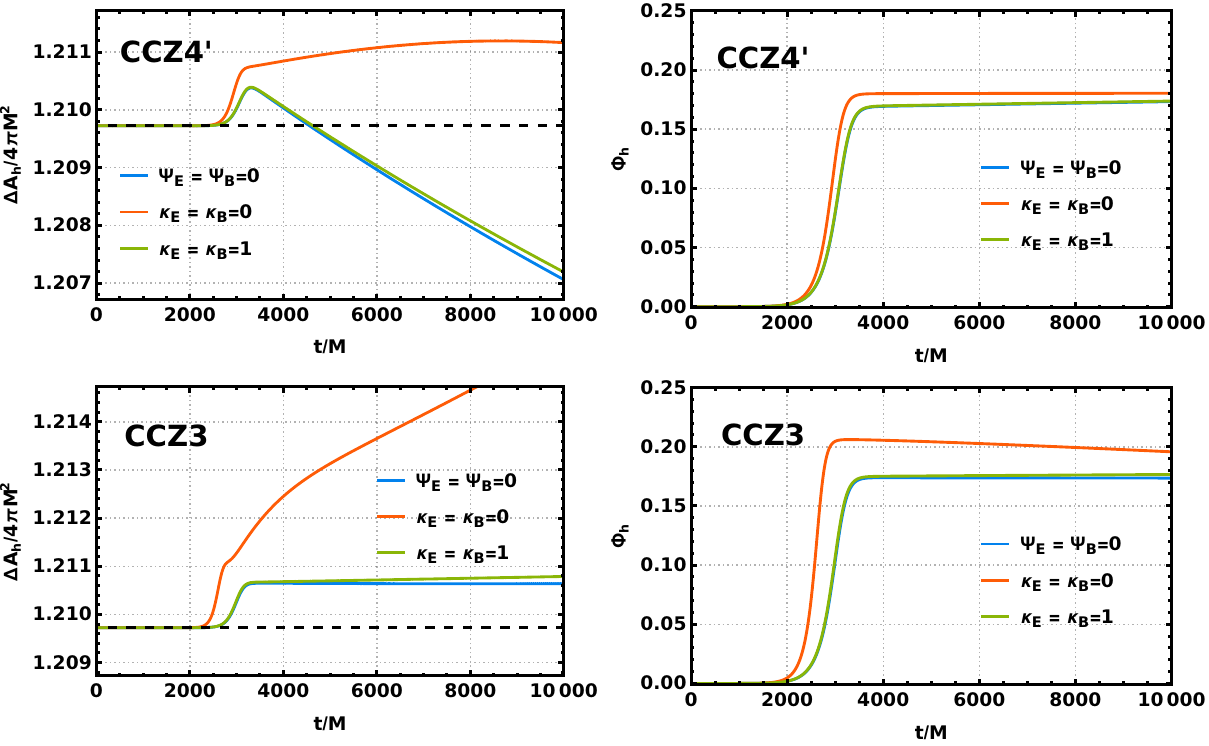}
\par\end{centering}
\caption{The evolution of the subtracted apparent horizon area $\Delta A_{h}$ (left panels), cf.\ Eq.\ \eqref{eq:Delta Ah}, and the horizon value of the scalar field $\Phi_{h}$ (right panels) for the spontaneous scalarization process, using the CCZ4' (upper panels) and CCZ3 (lower panels) formulations. The CCZ3 scheme is seen to exhibit the most accurate and robust results for the apparent horizon growth during the scalarization process, cf.\ in particular the blue line in the lower-left panel.}
\label{SBH_Phi}
\end{figure}

Finally, we study the comparison of the two new proposed schemes, CCZ4' and CCZ3, and considering different choices for the numerical formulation of the electromagnetic field constraints. To this end, we introduce the quantity
\begin{equation} \label{eq:Delta Ah}
\Delta A_{h}\left(t\right)=A_{h}^{\delta\Phi(p=10^{-4})}\left(t\right)-A_{h}^{\delta\Phi(p=0)}\left(t\right)+A_{h}^{\delta\Phi(p=0)}\left(0\right),
\end{equation}
which effectively subtracts the background noise $A_{h}^{\delta\Phi(p=0)}\left(t\right)$, as a way to better track the area of the apparent horizon during the scalarization process. Additionally, we compute the scalar field value $\Phi_{h}$ at the location of the apparent horizon. In Fig.\ \ref{SBH_Phi} we show the results of simulations using the two formulations. The CCZ4' scheme does not succeed to accurately simulate the evolution, at least insofar as the apparent horizon is concerned, which is seen not to converge (upper-left panel). On the other hand, the CCZ3 scheme is able to maintain numerical stability and yield convergent results, although we also learn that additional care is needed in the presence of constraints in the matter sector. In the case of the EMS system considered here, we see that different schemes for the evolution of the Gauss constraints lead to different outcomes, with the choice that results in highest accuracy and stability being the setting with non-propagating constraints, i.e.\ $\Psi_{E}=\Psi_{B}=0$ (cf.\ the blue curves in the lower panels of Fig.\ \ref{SBH_Phi}).

\section{Conclusions}
\label{Sec:Conc}

Our aim in this paper was to investigate the numerical stability of long-term black hole simulations within numerical relativity. We have identified a potential ``late-time instability'' in the evolution of static black holes, present already in the simplest context of the Schwarzschild spacetime, and which could not be resolved through various numerical manipulations. Drawing inspiration from the CCZ4 formulation, which is designed to propagate constraint violations as extra equations in order to enhance numerical stability, we have proposed several modified schemes aimed at mitigating this instability in black hole evolution.

Our numerical experiments with the evolution of a Schwarzschild black hole demonstrated that propagating the momentum constraint violation $Z_{i}$ without introducing a damping effect successfully alleviates late-time numerical instabilities. Based on this observation, we introduced two modified versions of the CCZ4 formulation, referred to here as CCZ4' and CCZ3, both of which enable stable long-term evolution, reaching times of order $10^5M$ in our simulations.

The advantage of the CCZ4 formulation over the BSSN approach is the implementation of constraint violations as propagating equations. Critical to the success of this method is the inclusion of appropriate damping factors. However, it has been appreciated that excessive damping may induce non-linear numerical instabilities, cf.\ Fig.\ \ref{Schw_variousScheme} and Refs.\ \cite{Alic:2013xsa,Sanchis-Gual:2014nha}. This observation underscores the need for a comprehensive assessment of different schemes in the implementation of damping factors in numerical relativity, which is precisely what we have endeavored to do in this paper.

We have found that this type of instabilities can be avoided provided that the damping effect on the momentum constraint violation, $Z_{i}$, is turned off, even with substantial damping applied to the Hamiltonian constraint violation $\Theta$ (cf.\ the upper-right panel of Fig.\ \ref{Schw_variousScheme} where $\kappa_{\Theta}=1$ and $\kappa_{\Gamma}=0$). As a result, both the CCZ4' and CCZ3 formulations were shown to significantly enhance the numerical stability of black hole evolution over long time scales. The importance of the momentum constraint versus the Hamiltonian constraint, at least insofar as the long-term evolution of black holes is concerned, is further made clear by the drawbacks of the CCZ0 scheme, which we introduced specifically for the purpose of comparison.

To further test our proposed formulations, we also studied their application to black hole spacetimes with matter fields, concretely the EMS model described in Section \ref{sec:Applications}. In these cases, we observed that the CCZ3 scheme is superior to the CCZ4', while also emphasizing that additional care is required in the presence of constraints in the matter sector of the theory. At least for the system we considered here, turning off the evolution of the matter constraints appears to be the appropriate choice in the context of physical instabilities with slow growth rates. It would be interesting to further test this numerical scheme in other set-ups with weak instabilities, such as rotating black holes undergoing superradiance.

\begin{acknowledgments}
We are grateful to Yiqian Chen, Cristian Joana and Shenkai Qiao for useful discussions and comments. SGS, GG and XW are supported by the NSFC (Grant Nos.\ 12250410250 and 12347133). SGS also acknowledges support from a Provincial Grant (Grant No.\ 2023QN10X389). PW is supported in part by the NSFC (Grant Nos.\ 12105191, 12275183, 12275184 and 11875196).
\end{acknowledgments}

\appendix

\section{Modifications to the $\Gamma$-driver gauge condition} \label{app:gamma}

\begin{figure}[t]
\begin{centering}
\includegraphics[scale=0.8]{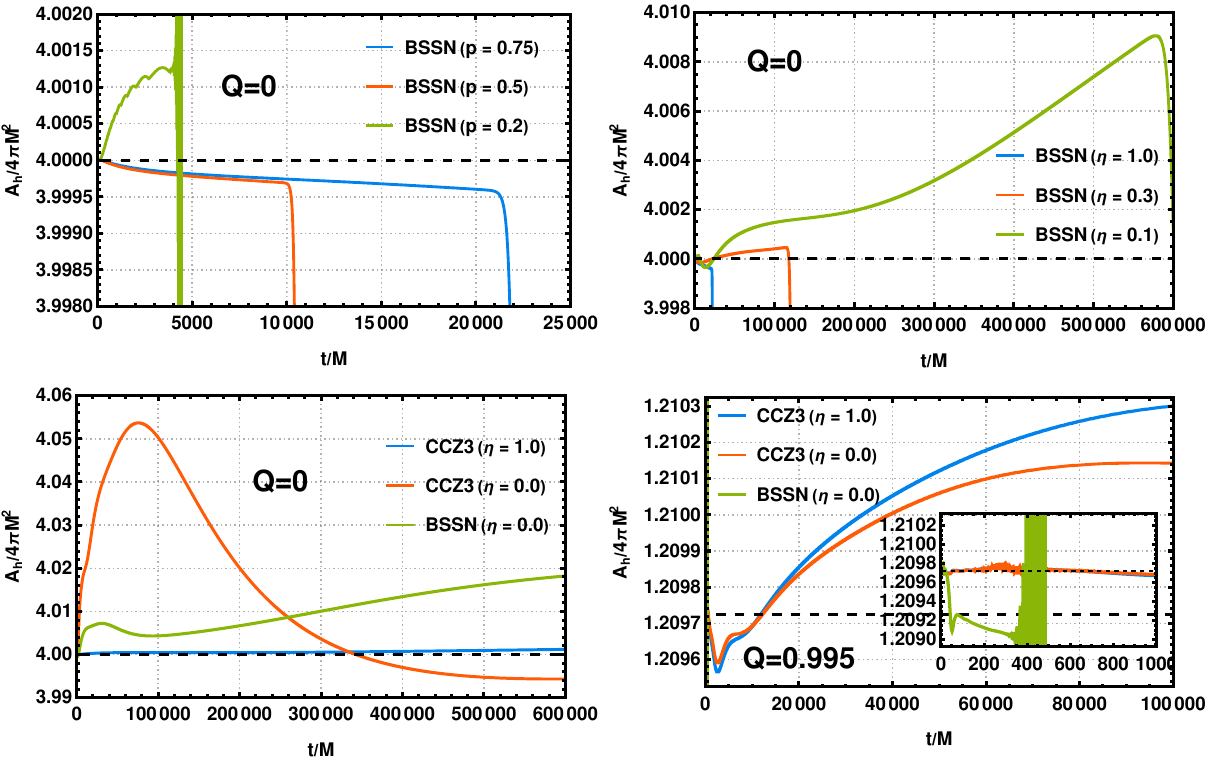}
\par\end{centering}
\caption{Evolution of Schwarzschild ($Q=0$) and RN ($Q=0.995$) black holes, as measured by the area of the apparent horizon, $A_h$, using various parameter settings in the $\Gamma$-driver gauge condition. The top two panels show the persistence of the long-term instability of a Schwarzschild black hole in the BSSN formulation under changes in the coefficients $p$ and $\eta$ in Eq.\ \eqref{eq:general gamma driver}. While the choice $\eta=0$ appears to cure the issue in the Schwarzschild case (bottom-left panel), this choice results in unstable simulations in the charged black hole set-up (bottom-right). In both cases, the CCZ3 scheme is seen to outperform BSSN.}
\label{fig:gamma}
\end{figure}

The $\Gamma$-driver gauge condition, Eq.\ \eqref{eq:gamma driver}, has been shown to be important in achieving stable long-term simulations of black holes \cite{Alcubierre:2002kk}. Throughout the main text we have used the standard choice $p=3/4$ and $\eta=1$ in the equation for the auxiliary vector,
\begin{equation} \label{eq:general gamma driver}
\partial_{\perp}C^{i}=p\partial_{\perp}\tilde{\Lambda}^{i}-\eta C^{i} .
\end{equation}
Both parameters $p$ and $\eta$ are however adjustable (see e.g.\ \cite{Yuwen:2024gcf} for a comparison of different values in another context). While $\eta$ plays the role of a damping parameter, $p$ may be associated to the speed of propagation of the shift vector, at least when the spacetime is approximately flat \cite{Alcubierre:2002kk}. In particular, $p=3/4$ means that the longitudinal component of $\beta^i$ propagates at the speed of light, which explains why this is a standard choice and also why larger values of $p$ may be prone to numerical instabilities, as indeed we have checked to be the case in our simulations. Moreover, we find that the long-term instability of the BSSN formulation, at least in the set-up of a Schwarzschild black hole, is also worsened by decreasing the parameter $p$ (cf.\ the top-left panel of Fig.\ \ref{fig:gamma}). Thus the choice $p=3/4$ appears to be roughly optimal within the context of long-term black hole evolution.

On the other hand, we have found that reducing the damping coefficient $\eta$ may result in improvements in the stability of the BSSN formulation when applied to a Schwarzschild black hole (top-right panel of Fig.\ \ref{fig:gamma}). Still, BSSN is seen to underperform when compared with the CCZ3 scheme (bottom-left panel), although the long-term instability appears to be resolved for the choice $\eta=0$. However, this seems to be peculiar to the Schwarzschild case, and indeed we can see that already in the set-up of a RN black hole the BSSN formulation with zero damping leads to instabilities in a relatively short timescale (bottom-right panel). Our simulations also suggest that the impact of changing $\eta$ in the CCZ3 scheme is unimportant.

\bibliographystyle{unsrturl}
\bibliography{ref}

\end{document}